\documentclass[twocolumn,aps,pra,superscriptaddress,showpacs,tightenlines]{revtex4}
\usepackage{amsmath}
\usepackage{amsfonts}
\usepackage{graphicx}
\usepackage{epsfig}
\usepackage{color}
\usepackage[colorlinks,citecolor=blue]{hyperref}
\begin{document}
\title{Mass sensing by quantum-criticality}
\author{Shang-Wu Bin}\affiliation{School of Physics, Huazhong University of Science and Technology, Wuhan, 430074, P. R. China}
\author{Xin-You L\"{u}}\email{xinyoulu@hust.edu.cn}
\affiliation{School of Physics, Huazhong University of Science and Technology, Wuhan, 430074, P. R. China}
\author{Tai-Shuang Yin}\affiliation{School of Physics, Huazhong University of Science and Technology, Wuhan, 430074, P. R. China}
\author{Gui-Lei Zhu}\affiliation{School of Physics, Huazhong University of Science and Technology, Wuhan, 430074, P. R. China}
\author{Qian Bin}\affiliation{School of Physics, Huazhong University of Science and Technology, Wuhan, 430074, P. R. China}
\author{Ying Wu}\email{yingwu2@126.com}
\affiliation{School of Physics, Huazhong University of Science and Technology, Wuhan, 430074, P. R. China}
\date{\today}
\begin{abstract}
Mass sensing connects the mass variation to a frequency shift of a mechanical oscillator, whose limitation is determined by its mechanical frequency resolution. Here we propose a method to enlarge a minute mechanical frequency shift, which is smaller than the linewidth of the mechanical oscillator, into a huge frequency shift of the normal mode. Explicitly, a frequency shift of about 20 Hz of the mechanical oscillator would be magnified to be a 1 MHz frequency shift in the normal mode, which increases it by five orders of magnitude. This enhancement relies on the sensitivity appearing near the quantum critical point of the electromechanical system. We show that a mechanical frequency shift of 1 Hz could be resolved with a mechanical resonance frequency $\omega_b = 11\times 2\pi $ MHz. Namely, an ultrasensitive mechanical mass sensor of the resolution $\Delta m /m \sim2\Delta \omega_b/\omega_b\sim 10^{-8}$ could be achieved. Our method has potential application in mass sensing and other techniques based on the frequency shift of a mechanical oscillator.
\end{abstract}
\maketitle
Nanomechanical oscillators take sensitivity on mass variation for the relation $\delta m=2m\frac{\delta\omega}{\omega}$ \cite{ekinci_nanoelectromechanical_2005}, where $m$ is the effective mass of nanomechanical oscillator and $\omega$ is the resonance frequency. Obviously, the mechanical oscillator with a higher frequency and lower mass would lead to a higher mass sensitivity in the same mechanical frequency shift. In a general way, frequency detection is implemented by monitoring the mechanical displacement which is driven by a microelectronic circuit with the frequency close to mechanical resonance frequency. The mechanical displacement corresponding to the frequency shift can be read out by the transducer-amplifier. Mass sensitivity has been improved from 7000 yg ($10^{-24}$ g) \cite{yang_zeptogram-scale_2006} to ~200 yg \cite{lassagne_ultrasensitive_2008, chiu_atomic-scale_2008}, even 1 yg \cite{chaste_nanomechanical_2012} in the last decades. In theory, this kind of mass sensor comes to a limit of sensitivity for the noise in mass adsorption process and the restriction on frequency monitor \cite{cleland_noise_2002,ref6}.

Optomechanical systems\cite{eichenfield2009optomechanical,RN_59_2014,RN_2003_209} features the optical force to displace mechanical oscillator. For a weak optomechanical coupling, a strong optical driving leads to optomechanically induced transparency (OMIT) \cite{RN_215,RN_pra_2010,RN_199_2011,RN_152,RN_13} and high-order sidebands \cite{RN_159,PA_97} even though it loses the optomechanical nonlinearity. When the optomechanical coupling is enhanced by other strategies \cite{PhysRevLett_119_053601,lu_squeezed_2015,RN_2,RN_2013_42,RN_2013_201,RN_2013_200}, quantum nature of the optomechanical system would be revealed by photon blockade \cite{RN_173,RN_143}, nonclassical states \cite{RN_143,yin_nonlinear_2017} and entanglement \cite{wang_macroscopic_2016,RN_177,RN_2013_208,RN_2013_206,RN_2013_207}. Recently, an all-optical method based on optomechanical systems comes up to weigh the mass of human DNA \cite{li_plasmon-assisted_2011,li_nonlinear_2012}, which avoids heating effect in the traditional mass sensor. This new method resolves the mechanical frequency shift from the movement of stokes field (one-order sideband) in the cavity output spectrum. However, both of the two methods rely on the resolution to the minimum value of the mechanical frequency shift. For the all-optical way, the limit of frequency resolution is the linewidth of mechanical oscillator, since the stokes field has the same linewidth with mechanical oscillator and a frequency shift smaller than the linewidth can not be resolved from the spectrum. To resolve a smaller mechanical frequency shift, this kind of mass sensor tends to adopt a smaller linewidth \cite{Jiang_2014}. But this would generally come along with the decrease of mechanical resonance frequency, which can not efficiently improve the precision of mass sensing. Therefore, one urgent question would be how to break through this limit \cite{cleland_noise_2002}.

We propose a method to magnify a minute mechanical frequency shift, which could be smaller than the linewidth of the mechanical oscillator, into an visible size on the normal mode. Thus we may concentrate on the operable normal mode to avoid sticking in the mire of resolution limit on mechanical frequency shift. Our proposal can be implemented in a hybrid electro-optomechanical system, in which the sensitivity is induced by quantum criticality \cite{RN_2,RN_2013_42,RN_2013_201,RN_2013_200,RN_2015_202,Zhu_2018,RN_2009_211}. This sensitivity to small mechanical frequency change can be achieved in the system with an optomechanical-like coupling. The improvement is very prominent that the magnitude of frequency shift would be magnified from 20 Hz in the mechanical oscillator to 1 MHz in the normal mode within the proper parameters. And because frequency shift of normal mode has an one-to-one correspondence with mechanical resonance frequency shift, we only need to focus on resolving the frequency shift of normal mode.  Fortunately, a strong coupling between normal mode and optical mode can be realized in our system, such that normal frequency shift would be simply exhibited in the cavity emission spectrum. A frequency shift of 1 Hz in the mechanical oscillator could be resolved with setting the mechanical frequency $\omega_b=11\times 2\pi$ MHz and linewidth $\Gamma_b=32\times 2\pi$ Hz. It would be hard to resolve in the all-optical way for 1 {\rm Hz} $\ll \Gamma_b$. Ultrahigh level of mass sensing can be reached to be $\Delta m /m=2\Delta \omega_b/\omega_b \sim 10^{-8}$ under this resolution of mechanical resonance frequency. This enhancement of sensitivity by quantum-criticality would not require a very strong coupling between the optical cavity and the mechanical oscillator, so we believe that it is achievable for future experiments.
\begin{figure}[htb]
\centerline{\includegraphics[width=8cm]{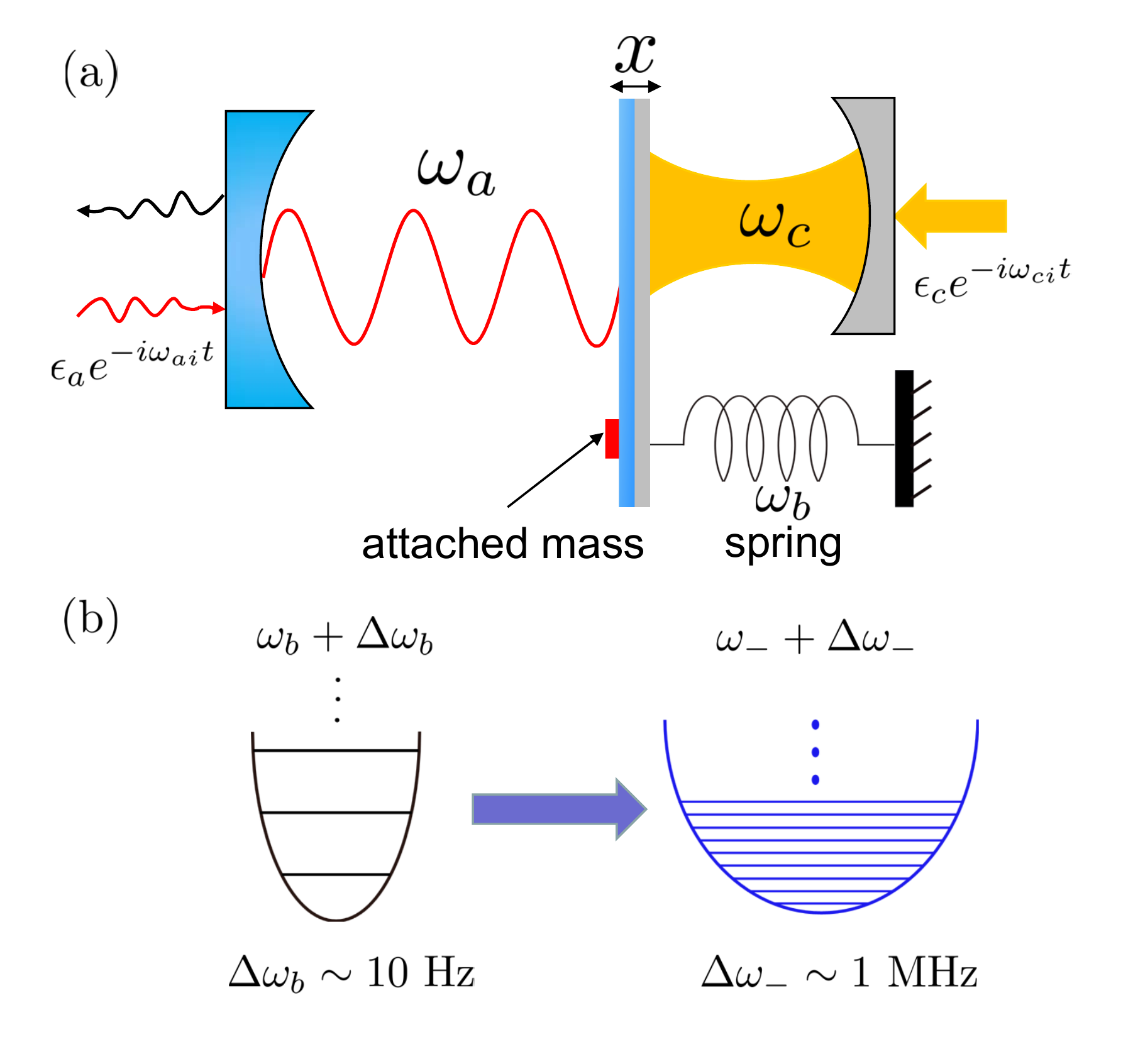}}
\caption{(Color online) (a) Hybrid elctro-optomechanical system for mass sensing. The mechanical oscillator(spring) couples with optical mode $\omega_a$ and microwave mode $\omega_c$, respectively. With strong driving $\epsilon_c$, microwave mode and mechanical oscillator generate the normal mode sensitive to mechanical frequency shift. The weak probe field $\epsilon_a$ is used for the visualization of ultrasmall mass change. (b) Energy level diagrams of the mechanical oscillator and the normal mode with $\omega_-$. $\Delta\omega_b$ is the frequency shift induced by small mass change. }
\label{fig1}
\end{figure}

As depicted in Fig. \ref{fig1}(a), here we consider the system consisting of an optomechanical system and a microwave resonator which is also coupled with the mechanical oscillator. The attached mass is located on the spring(mechanical oscillator) with resonance frequency $\omega_b$. We use $\omega_a$ ($\omega_c$) to denote the resonance frequency of optical mode(microwave mode). The linewidth of mechanical oscillator(microwave mode or optical mode) is $\Gamma_b$ ($\kappa_c$ or $\kappa_a$), respectively. The microwave mode is driven by a strong field with amplitude $\epsilon_c=\sqrt{2 P_c \kappa_c/\hbar \omega_{ci}}$, where $ P_c$ is the power and $\omega_{ci}$ is the frequency. In the rotating frame of $\omega_{ci}$, the Hamiltonian reads
\begin{align}
H_{{\rm OMS}}/\hbar & = \delta_c c^\dag c + \omega_a a^\dag a +\omega_b b^\dag b+g_a a^\dag a (b^\dag+b)\notag\\
& +g_c c^\dag c (b^\dag +b) + i \varepsilon_c (c^\dag -c) ,
\end{align}
where $a$ ($c$ or $b$) represents the annihilation operator of the optical cavity (microwave or mechanical mode). $\delta_c=\omega_c-\omega_{ci}$ is the detuning of microwave mode with respect to the strong driving field, and $g_a$ ($g_c$) stands for the optical (microwave) cavity coupling with mechanical oscillator, respectively. Under the strong driving field, there is a standard linearization procedure \cite{RN_177,RN_175,RN_174} to give out the Hamiltonian
\begin{align}
H^\prime_{{\rm OMS}}/\hbar & = \Delta_c c^\dag c + \tilde{\omega}_a a^\dag a +\omega_b b^\dag b+g_a a^\dag a (b^\dag+b)\notag\\
& - G(c^\dag +c)(b^\dag +b) ,
\end{align}
with
\begin{align}
G &=g_c\sqrt{\frac{2P_c\kappa_c}{\hbar (\omega_c-\delta_c)(\kappa_c^2+\Delta_c^2)}}  ,\\
\Delta_c &= \delta_c - \frac{4g_c^2P_c\kappa_c}{\hbar \omega_b(\omega_c-\delta_c)(\kappa_c^2+\Delta_c^2)}  ,\\
\tilde{\omega}_a &= \omega_a - \frac{4g_ag_cP_c\kappa_c}{\hbar \omega_b(\omega_c-\delta_c)(\kappa_c^2+\Delta_c^2)}  ,
\end{align}
where $G$ is the linearized electromechanical coupling, $\Delta_c$ is the effective microwave detuning and $\tilde{\omega}_c$ is the redefined optical frequency, respectively.
The reason for the amplification effect would be explicit after diagonalizing the electromechanical subsystem via the Bogoliubov transformation $\hat R=M\hat B$, where $\hat R^T=(c,c^\dag,b,b^\dag)$ and $\hat B^T=(B_-,B^\dag_-,B_+,B^\dag_+)$. $M$ is the transform matrix, which is the same in Ref. \cite{RN_2}.

The Hamiltonian is transformed into
\begin{align}
H^\prime _{{\rm TOMS}}/\hbar & = \omega_- B_-^\dag B_- + \omega_+ B^\dag_+B_+ + \tilde{\omega}_a a^\dag a \notag\\ &+g_-a^\dag a(B^\dag_-+B_-) +g_+a^\dag a (B^\dag_+ +B_+) ,
\end{align}
with

\begin{align}
\omega^2_{\pm}=\frac{1}{2}(\Delta_c^2+\omega_b^2\pm\sqrt{(\omega_b^2-\Delta_c^2)^2+16G^2\Delta_c\omega_b}) ,
\label{eq7}
\end{align}
and
\begin{align}
g_{\pm}=\pm g_a\sqrt{\omega_b(1\pm\cos{2\theta})/2\omega_{\pm}} ,
\end{align}
are the redefined coupling between the effective normal mode and the mechanical oscillator. The angle $\theta$ can be obtained from
\begin{align}
\tan2\theta=\frac{4G\sqrt{\Delta_c\omega_b}}{\Delta_c^2-\omega_b^2} .
\end{align}
We would like to give a physical explanation for the transformation \cite{RN_2,RN_2013_42,RN_2013_201,RN_2013_200,RN_2015_202}. The microwave mode and mechanical mode with coordinate-coordinate coupling intervene with each other to generate two normal modes with resonance frequency $\omega_-$ and $\omega_+$, respectively, where $\omega_-$ is smaller than $\omega_+$. Adjusting the coupling strength to proper value, $\omega_-$ would approach to zero even negative, which implies the $\omega_-$ normal mode changes from a harmonic oscillator to unstable free particle. The condition $\omega_-=0$ is defined to be the quantum critical point. In this work, we would only focus on the stable regime. When the system is close to the critical point but in the stable case, the high sensitivity to mechanical frequency shift appears. As shown in Fig. \ref{fig1}(b), a normal normal mode with such a low resonance frequency makes itself very sensitive to the change of system parameters, in which the mechanical resonance frequency are our interest. We would like to point out that the stability of the parameters is required for the very sensitivity to mechanical frequency shift.
\begin{figure}[htb]
\centerline{\includegraphics[width=8cm]{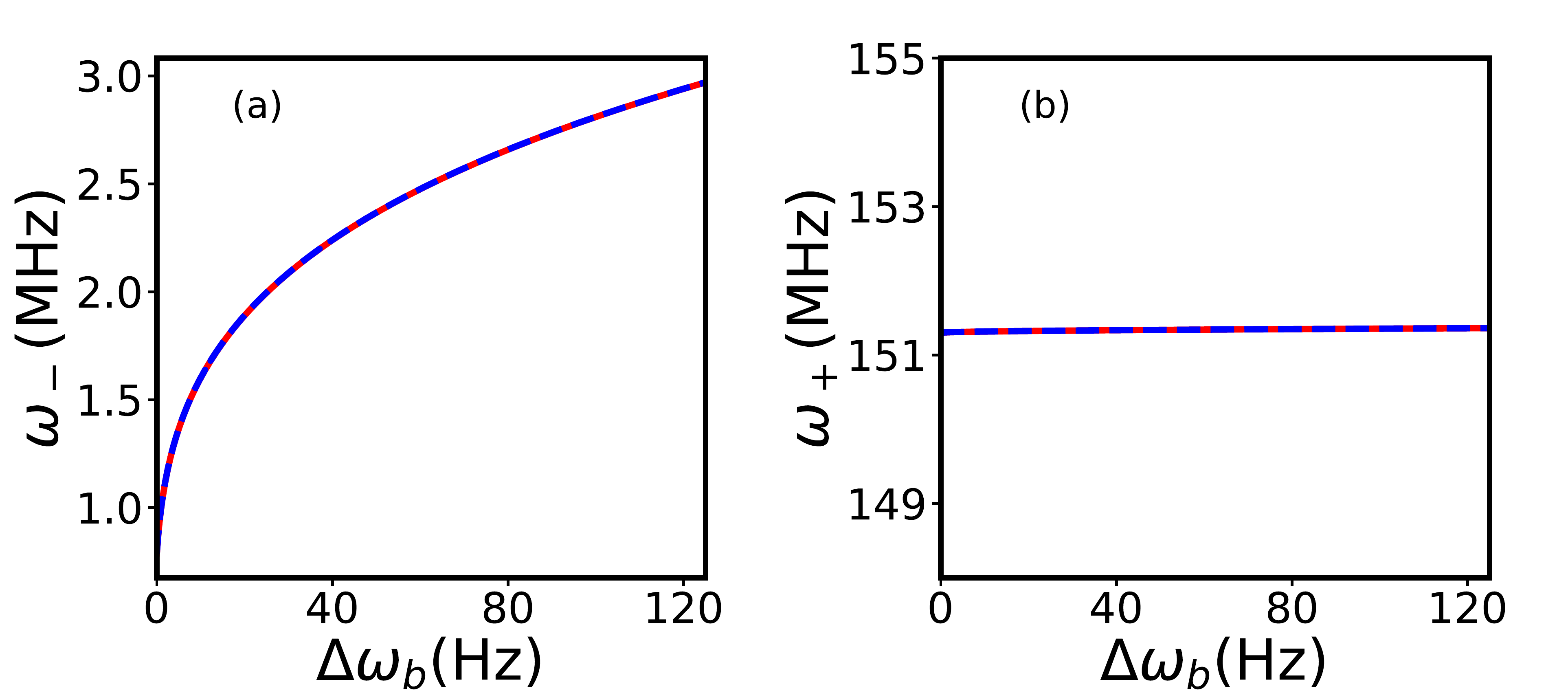}}
\caption{(Color online) (a) The resonance frequency $\omega_-$ of normal mode. (b) The resonance frequency $\omega_+$ of normal mode. Blue dash (red solid) curve corresponds to numerical (analytical) calculation for normal resonance frequency. The parameter are $\kappa_a/2\pi=10 \kappa_c/2\pi =0.02 \omega_b$, $\delta_c\approx201.883$ MHz, $ P_c=0.3$ $\mu$W. $\Delta \omega_b=\omega^\prime_b-\omega_b$, and $\omega^\prime_b$ is the mechanical frequency after adsorbing tiny mass.}
\label{fig2}
\end{figure}

\begin{figure}[htb]
\centerline{\includegraphics[width=8cm]{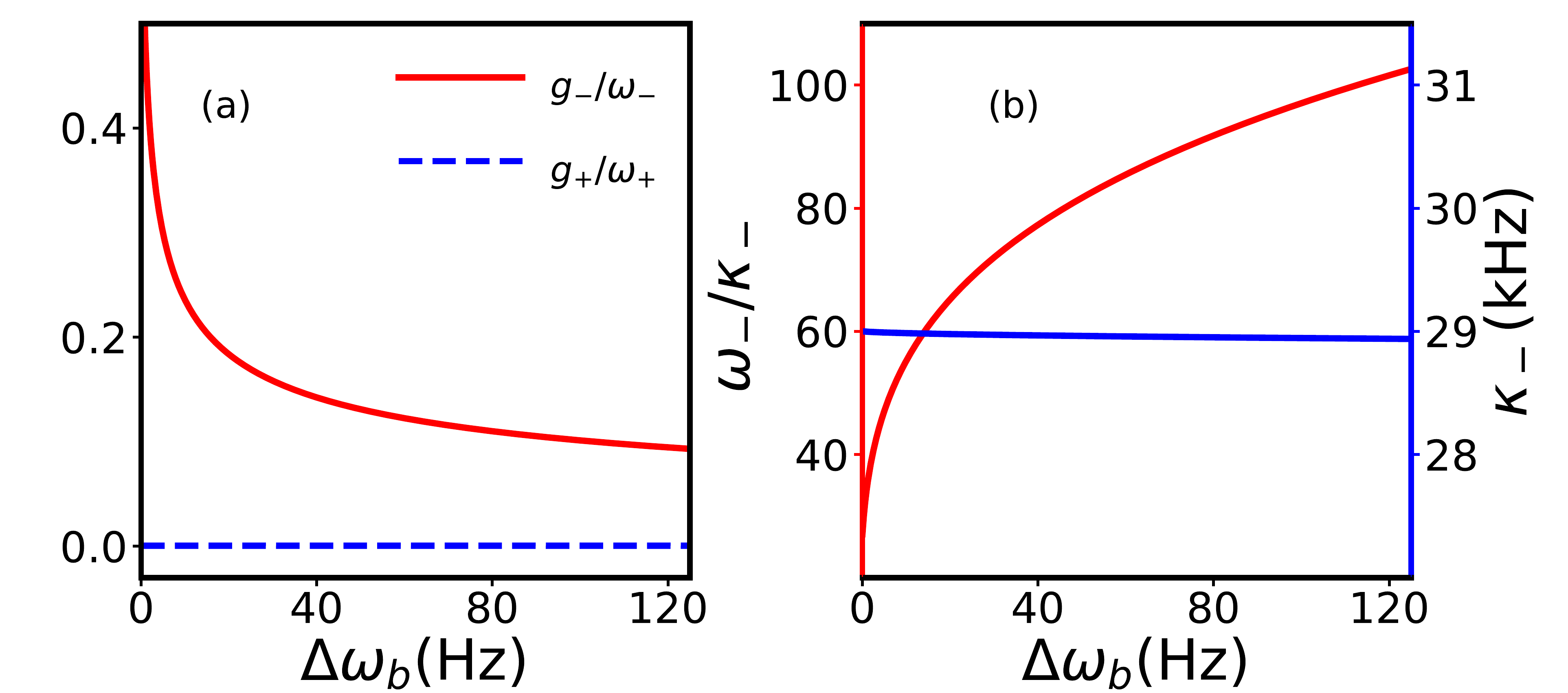}}
\caption{(Color online) (a) Coupling between the normal modes and optical mode. (b) Frequency-to-decay ratio and linewidth of the normal mode $\omega_-$. Parameters are the same with Fig. 2.}
\label{fig3}
\end{figure}

Figure. \ref{fig2} shows that the frequencies of two normal modes change with the mechanical frequency shift according to Eq. (\ref{eq7}) and numerical calculation, respectively. The numerical calculation obtains $\omega_-$ and $\omega_+$ by diagonalizing the coefficient matrix of Hisenberg-Langevin equations of mechanical mode $b$ and microwave mode $c$. Then $\kappa_-$ and $\kappa_+$ corresponding to the linewidth of normal mode $\omega_-$ and $\omega_+$, respectively, come up when the decay $\Gamma_b$ ($\kappa_c$) of mechanical oscillator (microwave cavity) is considered \cite{RN_2,RN_2013_42,RN_2013_201,RN_2013_200,RN_2015_202}. In this work, we set the main parameters as ($\omega_b$, $\omega_c$, $\omega_a$, $\Gamma_b$, $g_c$, $g_a$)=(11 MHz, $7.5$ GHz, $12.9$ GHz, 32 Hz, 20 kHz, 20 kHz)$\times2\pi$ \cite{RN_199_2011,RN_59_2014,RN_2008_212}. Under those conditions, We also note that $\kappa_-$ and $\kappa_+$ are mainly determined by $\kappa_c$ for $\Gamma_b\ll\kappa_c$. An explicit form from Ref. \cite{RN_2015_202} is $\kappa_-=\Gamma_b( M_{31} +  M_{32})^2+\kappa_c( M_{11}^2- M_{12}^2)$, where $ M_{ij}$ is the matrix element of transform matrix $M$. It is clearly shown in Fig. \ref{fig2}(a) that there is an enlargement of mechanical frequency shift. Specifically, a tiny mechanical resonance frequency shift about 20 Hz induces 1 MHz frequency shift in normal mode frequency. Actually, the frequencies of both two normal modes have an amplification effect of mechanical frequency shift, while $\omega_+$ alters little with respect to itself. In addition, Fig. \ref{fig3}(a) exhibits that the $\omega_+$ mode leads to a very weak optomechanical nonlinearity ($g_+/\omega_+\approx 0$) for an occupying very high energy level. Therefore, the $\omega_+$ mode can be ignored safety in the weak probe field regime ($\epsilon_a\ll\kappa_a$). On the contrary, the $\omega_-$ mode possesses a strong coupling with the optical mode. In a simple way to show the frequency shift of normal mode, we may concentrate on the cavity emission spectrum which exhibits the sidebands generated by the interference between optical mode and normal mode. In other words, we tailor the Hamiltonian as
\begin{align}
H^\prime_{{\rm OM}}/\hbar & = \Delta_a a^\dag a +\omega_- B_-^\dag B_- +g_- a^\dag a (B_-^\dag+B_-)\notag\\
& + i\epsilon_a(a^\dag -a) ,
\end{align}
where $\Delta_a=\tilde{\omega}_a-\omega_L$ is the detuning between the probe field and optical mode.

Recently, an all-optical technique is applied in mass sensing to attain a high resolution \cite{li_plasmon-assisted_2011,li_nonlinear_2012}. Specifically, the mechanical frequency shift is resolved from the movement of stokes field in the cavity emission spectrum.  In those schemes, the mechanical oscillator is suggested to be cooled down and operated in a ultrahigh vacuum. Thus, the limit of resolution is determined by the full width half maximum (FWHM) of the peak of stokes field, i.e., the linewidth of mechancal oscillator. In our proposal, a similar restriction that the resolution of normal frequency shift is $\kappa_-$ comes if we wish to resolve frequency shift from the emission spectrum. $\kappa_-$ and the corresponding bath thermal occupancies $n_-$ can be calculated from the diagonalization process of elelctromechanical system($\omega_b$ and $\omega_c$). We consider that the temperature of mechanical oscillator is very low, and the experiment is implemented in a ultrahigh vacuum. In this case, $n_-\ll 1$ is a suitable  assumption \cite{RN_2013_42,RN_2013_201,RN_2013_200,RN_2015_202}. For the range of mechanical frequency shift in which we are interested, $\kappa_-$ alters little and satisfies the sideband resolution condition ($\omega_->\kappa_-$) from Fig. \ref{fig3}(b).

As discussed above, the master equation of system is expressed as
\begin{align}
\dot\rho = &-i[H^\prime_{{\rm OM}},\rho] + (n_-+1)\kappa_-\mathcal{D}[B_-]\rho +n_-\kappa_-\mathcal{D}[B^\dag_-]\notag\\
& +\kappa_a\mathcal{D}[a]\rho .
\end{align}
$\mathcal{D}[o]\rho=o\rho o^\dag-(o^\dag o\rho+\rho o^\dag o)/2$ is the standard dissipator in Lindblad form.
To reveal the effect of the mechanical frequency shift, We define the cavity emission spectrum
\begin{align}
S(\omega)=\int^\infty_{-\infty}\mathrm{d}t\mathrm{e}^{i\omega t} \kappa_a\langle a^\dag (t)a(0) \rangle .
\end{align}
In the limit $\kappa_a\gg\kappa_-$, the spectrum satisfies \cite{RN_143,RN_168}
\begin{align}
S(\omega)\sim \sum^\infty_{m,n=0}\frac{\kappa_-}{[\frac{(m+n)\kappa_-}{2}]^2+[\omega-(m-n)\omega_-]^2} .
\label{eq13}
\end{align}
Eq. (\ref{eq13}) indicates that the multi-sidebands are generated by the displaced number states of $\omega_-$ mode. The high-order sidebands are submerged by the factor $(m+n)\kappa_-$.

\begin{figure}[htb]
\centerline{\includegraphics[width=8cm]{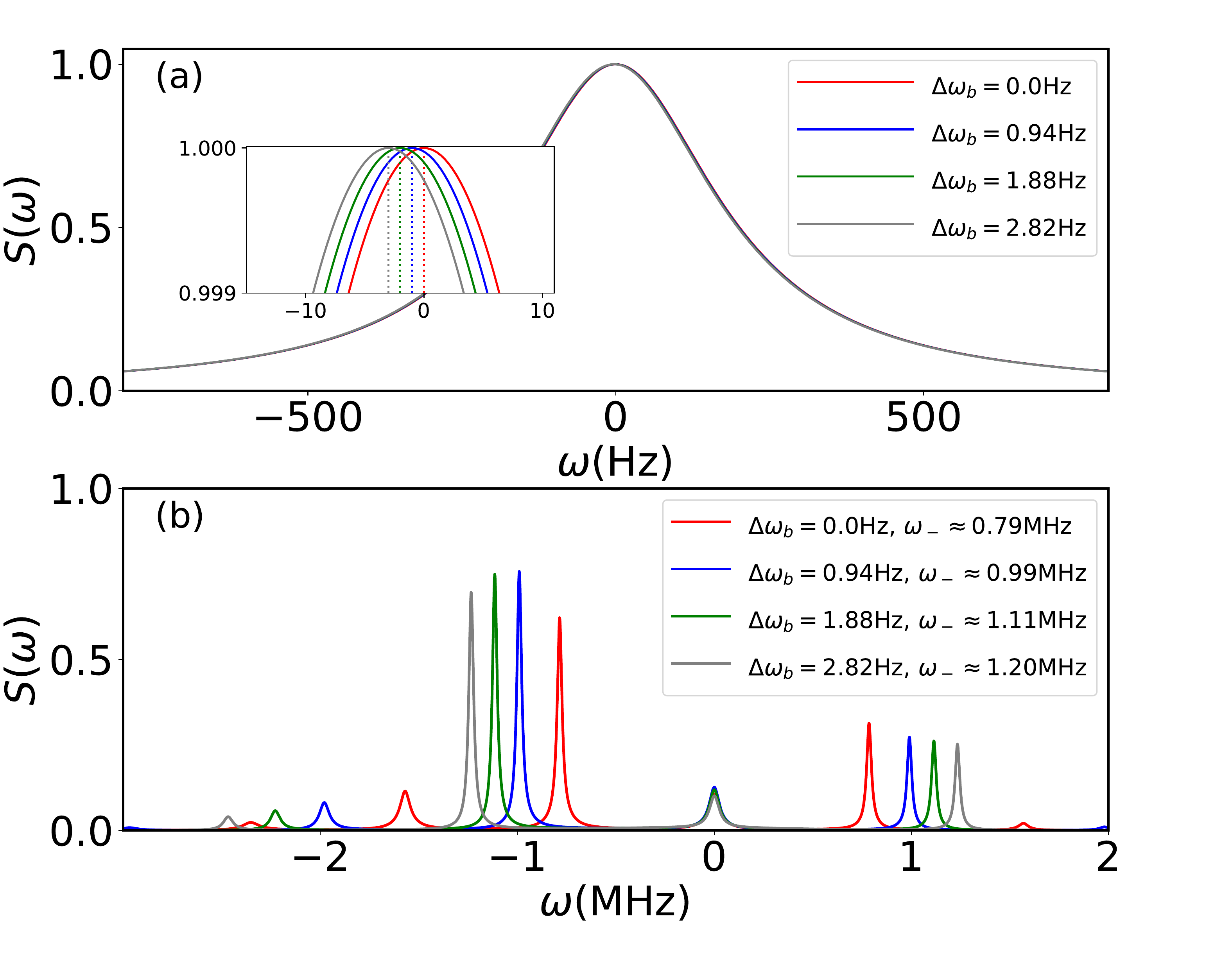}}
\caption{(Color online) Cavity Spectrum. (a) Spectrum without quantum-criticality enhancement. Inset: Enlargement of the peaks in the main figure. (b) Spectrum with the quantum-criticality enhancement. We pick up four points from Fig \ref{fig2}(a). $\Delta_a=0$, $\epsilon_a =0.1 \kappa_a$ and $n_-=0.1$, other parameters are the same with Fig. \ref{fig2}.}
\label{fig4}
\end{figure}
The cavity emission spectrums corresponding to the frequency shift of normal mode are shown in Fig. \ref{fig4}(b). To compare, we give out the spectrum of mechanical oscillator with Lorentz type in Fig. \ref{fig4}(a). The mechanical oscillator has the same resonance frequency and linewidth as the one of our system, i.e, $\Gamma_b =32\times 2\pi$ Hz and $\omega_b = 1.1\times10^7\times 2\pi$ Hz, while frequency shift of 1.0 Hz is very hard to resolve. In Fig. \ref{fig4}(b), we pick up four points from Fig. \ref{fig2}(a) to show their corresponding cavity emission spectrum. These mechanical frequencies increase with 1.0 Hz. The result meets the expectation as we discuss above. Ultrahigh mechanical frequency shift $\Delta\omega_b/\omega_b = 1/1.1\times 10^7 \times 2\pi\approx 10^{-8}$ can be resolved. The minute mechanical frequency shift becomes obvious in the cavity emission spectrum. We would like to emphasize that this 1 Hz mechanical frequency shift is chose to be resolved for the condition $\delta \omega_- > \kappa_-$. This requirement is only necessary for such simple spectral detection. One may reach a higher precision for other measure way which is immune to $\kappa_-$ \cite{lin_mass_2017}.

In summary, we propose a method to enlarge a minute mechanical frequency shift to be observable, which relies on the very sensitivity near the quantum-criticality point. A frequency shift of mechanical oscillator well below the mechanical linewidth could be resolved, which is a break through of the previous works. Our proposal can also operate in a all-optical condition to avoid the heating effect in the traditional nanomechanical systems. The level of small frequency shift about 20 Hz would be enlarged to around 1 MHz in the normal mode. In a simple spectral way to monitoring the frequency shift of normal mode, the resolution is limited to the effective decay of the normal mode. We note that this resolution would require a cryogenically cooled and ultrahigh vacuum apparatus. As we have shown above, ultrasmall mass resolution $\Delta m/m \sim 10^{-8}$ is straightforward. For example, if we choose the mass of mechanical oscillator to be $4.8\times10^{-14}$ kg \cite{RN_199_2011,RN_59_2014}, resolution of mass sensing would be up to 1.4 $ag$ ($10^{-18}$g). More importantly, this quantum-criticality possesses great potential to explore the extreme high precision of mechanical frequency shift. This may be applied in various sensors based on the mechanical frequency detection. Experimentally, this mass sensor may be implemented by the general hybrid elctro-optomechanical system as we discussed above or the ring resonator with two degenerate optical modes coupled to a mechanical mode \cite{PhysRevLett_117_110802,Jing_18,chen_yuan,RN_2009_211}, respectively. The quantum-criticality can be realized without the strong optomechanical coupling. For the excellent resolution of mechanical frequency shift and experimental achievement in the future, our scheme possesses great potential development in mass sensing.

Shang-Wu Bin thanks Andreas Nunnenkamp for helpful suggestions. This work is supported by the National Key Research and Development Program of China (2016YFA0301203); National Natural Science Foundation of China (NSFC) (11374116, 11375067, 11574104).
\bibliographystyle{unsrt}
\bibliography{mass_sensor2,mass_sensor3}
\end{document}